\documentclass[10pt, conference, a4paper]{IEEEtran}

\usepackage{cite}
\usepackage{amsmath,amssymb,amsfonts}

\usepackage{subcaption}

\usepackage{amsthm}
\usepackage{epsfig}
\usepackage{color}
\usepackage{lettrine}
\usepackage{float}
\usepackage{lipsum}
\usepackage{bm}
\usepackage{mathtools}
\usepackage{graphics}
\usepackage{bbm}
\usepackage{etoolbox}
\usepackage{suffix}
\usepackage{booktabs}
\usepackage[T1]{fontenc}
\usepackage[colorlinks]{hyperref} 
\usepackage[nameinlink,noabbrev]{cleveref}
\usepackage{textcomp}
\usepackage{xcolor}
\usepackage{stackengine}
\usepackage[ruled,norelsize]{algorithm2e}
\usepackage[noend]{algpseudocode}

\def \treq {\stackrel{\tiny \Delta}{=}}

\newcommand{\q}{\ensuremath{\mathbf q}}
\newcommand{\Q}{\ensuremath{\mathbf Q}}
\newcommand{\Qi}{\ensuremath{{Q}}^{-1}}
\newcommand{\e}{\mathbf{e}}
\newcommand{\E}{\ensuremath{\mathbb E}}

\renewcommand{\L}{\ensuremath{\mathbf L}}

\renewcommand{\O}{\ensuremath{\mathcal  O}}

\newcommand{\Prob}[1]{(\textbf{P}{#1})}
\newcommand{\cst}[1]{\textbf{C}{#1}}

\newcommand{\prop}{\textbf{JTBD}}
\newcommand{\rd}{\textbf{BDFT}}
\newcommand{\td}{\textbf{TDFB}}
\newcommand{\east}{\ensuremath{\mathrm{EAST}}}

\newtheorem{lemma}{Lemma}
\DeclareMathOperator*{\argmax}{arg\,max} 

\def \treq {\stackrel{\tiny \Delta}{=}}

\makeatletter
\newcommand{\removelatexerror}{\let\@latex@error\@gobble}
\makeatother

\usepackage{hyperref} 
\hypersetup{
    colorlinks=true,
    citecolor=blue}
    
\begin{document}

\title{Secure Short-Packet Transmission with Aerial Relaying: Blocklength and Trajectory Co-Design}

\author{\IEEEauthorblockN{Milad Tatar Mamaghani\IEEEauthorrefmark{1}, Xiangyun Zhou\IEEEauthorrefmark{1}, Nan Yang\IEEEauthorrefmark{1}, and
A. Lee Swindlehurst\IEEEauthorrefmark{2}}
\IEEEauthorblockA{
\IEEEauthorrefmark{1}School of Engineering, Australian National University, Canberra, ACT 2601, Australia\\
\IEEEauthorrefmark{2}Henry Samueli School of Engineering, University of California, Irvine, CA 92697, USA}
Email:~\href{mailto:milad.tatarmamaghani@anu.edu.au}{\textcolor{black}{milad.tatarmamaghani@anu.edu.au}}, \href{mailto:xiangyun.zhou@anu.edu.au}{\textcolor{black}{xiangyun.zhou@anu.edu.au}}, \href{mailto:nan.yang@anu.edu.au}{\textcolor{black}{nan.yang@anu.edu.au}},   \href{mailto:swindle@uci.edu}{\textcolor{black}{swindle@uci.edu}}
\thanks{This work was supported by the Australian Research Council’s Discovery Projects funding scheme (project number DP220101318).}}

\IEEEoverridecommandlockouts\IEEEpubid{\begin{minipage}{\textwidth}\ \\[75pt] \centering
   \copyright 2023 IEEE. Personal use of this material is permitted. Permission from IEEE must be obtained for all other uses, in any current or future media, including reprinting/republishing this material for advertising or promotional purposes, creating new collective works, for resale or redistribution to servers or lists, or reuse of any copyrighted component of this work in other works. 
\end{minipage}} 

\maketitle

\begin{abstract}
In this paper, we propose a secure short-packet communication (SPC) system involving an unmanned aerial vehicle (UAV)-aided relay in the presence of a terrestrial passive eavesdropper. The considered system, which is applicable to various next-generation Internet-of-Things (IoT) networks, exploits a UAV as a mobile relay, facilitating the reliable and secure exchange of intermittent short packets between a pair of remote IoT devices with strict latency. Our objective is to improve the overall secrecy throughput performance of the system by carefully designing key parameters such as the coding blocklengths and the UAV trajectory. However, this inherently poses a challenging optimization problem that is difficult to solve optimally. To address the issue, we propose a low-complexity algorithm inspired by the block successive convex approximation approach, where we divide the original problem into two subproblems and solve them alternately until convergence. Numerical results demonstrate that the proposed design achieves significant performance improvements  relative to other benchmarks, and offer valuable insights into determining appropriate coding blocklengths and UAV trajectory.
\end{abstract}


\vspace*{5mm}
\section{Introduction}


Short-packet communication (SPC) is a critical component of emerging beyond-5G (B5G) wireless networks and Internet of Things (IoT) applications, where devices need to exchange short-packet data to fulfill low-latency and low-cost communication requirements. On the other hand, reduced channel coding gain associated with short-packet transmission poses a major hurdle to communication reliability in SPC. In addition, security issues such as eavesdropping in SPC-IoT networks are more pronounced due to confidential and sensitive SPC data that IoT networks frequently need to share in an open wireless environment, particularly for mission-critical scenarios  \cite{Durisi2016, Feng2021, Shirvanimoghaddam2019}. Physical-layer security (PLS) technology is a promising candidate for safeguarding SPC. PLS techniques exploit the physical-layer characteristics of wireless channels or smart signaling for communication secrecy through wiretap coding, without suffering from the high complexity of resource-demanding key-based security methods \cite{Yang2015}.


Nevertheless, conventional PLS techniques adopt wiretap codes based on the assumption of infinite blocklengths. Substantial studies have thus far been conducted to secure different wireless networks working in the infinite blocklength regime (e.g., see \cite{TatarMamaghani2018, TatarMamaghani2020} and references therein).  However, SPC generally involves the transmission of short packets on the order of tens of bytes as opposed to several kilobytes in conventional wireless systems. Consequently, PLS-based designs need to be meticulously revisited, as adopting the so-called \textit{Secrecy Capacity}, i.e,  a typical performance metric for conventional PLS systems \cite{Poor2017},  is no longer applicable for systems operating under SPC due to the finite blocklength assumption. In light of this, \cite{Yang2019} fundamentally studied the attainable secrecy rate (SR) in a wiretap channel with specific reliability and secrecy requirements under the finite blocklength assumption. The authors in \cite{Zheng2020} and \cite{Feng2022}, developed PLS schemes for SPC assuming fading channels.  The work in \cite{Wang2019e} holistically investigated the performance of secure SPC in a mission-critical IoT system with an external adversary.  Nevertheless,  these studies have only considered system designs that involve stationary communication nodes and adopted a fixed number of information bits per short-packet transmission.  Consequently, the approaches developed in \cite{Zheng2020, Feng2022, Wang2019e} may not work well in highly dynamic scenarios or when the number of information bits generated for transmission by IoT devices varies.

Unmanned aerial vehicles (UAVs), with their relatively rapid on-demand deployment, low-cost maintenance, and maneuverability, can potentially be used in  a myriad of wireless applications ranging from serving as an aerial base station (BS) or mobile relay for remote sensing and real-time monitoring in IoT networks \cite{Wu2021}. Accordingly, it has been anticipated that integrating UAV technology into forthcoming B5G IoT systems is a cost-effective promising solution with substantial benefits. Due to their operation at relatively high altitudes, UAVs are particularly helpful in improving coverage by reducing signal attenuation in wireless links caused by blockage or shadowing. As a result, UAV-empowered wireless IoT systems, if properly designed, would require less power for transmissions. This is specifically beneficial for energy-hungry IoT networks since it expands their operational  lifetime. On the other hand, UAV-IoT scenarios are more susceptible to eavesdropping attacks due to line-of-sight (LoS)-dominant air-ground (AG) links \cite{Wang2019}. Thus, securing such systems presents significant challenges, and of course,  requires further research.



To address the above-mentioned challenges, in this work we propose a secure UAV-aided relaying scheme with SPC, where sensitive short packets need to be periodically transmitted from a remote IoT device to a designated receiver with a stringent latency requirement while combating passive eavesdropping.  We formulate a new optimization problem for the considered UAV-aided SPC system under security, reliability, latency, and mobility constraints. The formulated problem is nonconvex, and hence challenging to solve optimally.  To tackle the challenging nonconvex problem, we apply the  block successive convex approximation (BSCA) approach to iteratively solve a sequence of convex subproblems:  coding blocklength optimization and UAV trajectory design. We then propose a low-complexity algorithm combining these solutions to optimize the system performance. We conduct simulations to draw some useful insights into the performance of the proposed joint blocklength optimization and trajectory design, and highlight its secrecy performance advantage compared to other competitive benchmarks.

\vspace*{5mm}
\section{System Model and Assumptions}\label{sec:sysmodel}

\begin{figure}[!t]
\centering
\includegraphics[width=\columnwidth]{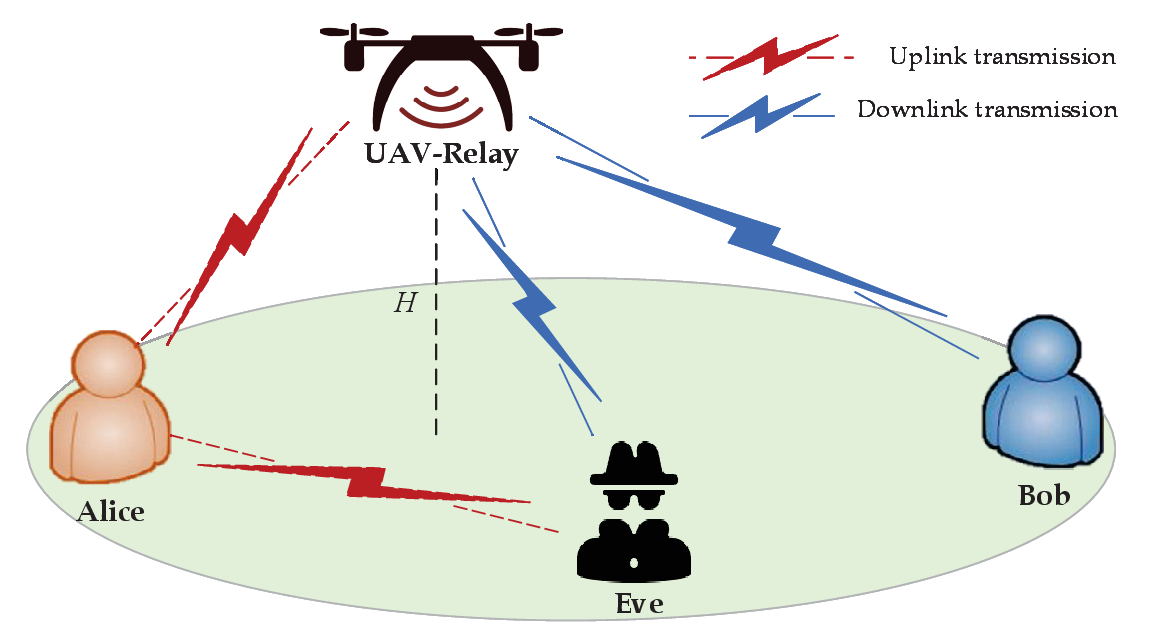}
\caption{ System model for secure UAV-SPC relaying.}
\label{fig1}
\end{figure}
As illustrated in Fig. \ref{fig1}, we consider a UAV-aided IoT communication system with secure SPC, wherein a source (Alice) periodically sends short packets containing confidential information to a designated remote destination (Bob) via a mobile UAV-Relay (UR),  while a passive eavesdropper (Eve) attempts to overhear the ongoing confidential transmissions. In practice, Alice periodically generates short-packet sensitive information from the environment and, if feasible, immediately transmits it to Bob with a stringent latency requirement. We assume that a packet is generated at the beginning of each \textit{timeslot} and the period of each timeslot is $\delta_t$. We also assume that Alice generates and sends a varying amount of information bits during each timeslot to handle different tasks such as monitoring, controlling, or sensing.

In this work, we assume that the direct link between Alice and Bob is absent due to long distance or blockages, and hence a UAV-mounted relay is employed to facilitate the end-to-end SPC. Moreover, we assume that  all the communication nodes are equipped with a single antenna, as commonly considered for  low-cost and resource-constrained IoT devices (see \cite{TatarMamaghani2021} and references therein). In addition, the mobile relaying strategy adopted by the UR is assumed to be the decode-and-forward (DF) protocol with time division duplexing (TDD), operating in a shared bandwidth $W$ for both reception and transmission. We assume that the UR-aided DF relaying for SPC occurs at the beginning of each timeslot $\delta_t$, which consists of two phases. In the first phase (i.e., uplink transmission) in timeslot $n$, Alice generates a short packet, containing sensitive information,  and transmits the packet with fixed  power $p_a$ to the UR over $l_u[n]$ channel uses, where $n=\{1, 2, \cdots\}$ denotes the index of each timeslot, and then the UR decodes the received signal to obtain the transmitted confidential message. In the second phase (i.e., downlink transmission) in timeslot $n$, the UR encodes the obtained message from the first phase with a different codebook for security purposes, forwarding the result to Bob over $l_d[n]$  channel uses with fixed relaying power $p_r$ and Bob retrieves the original confidential information.  While  Eve  wiretaps the ongoing transmissions in two phases to obtain confidential data, since the signals from both Alice and the UR are encoded with different codebooks, she cannot exploit a diversity combining strategy to improve her reception and  pose stronger security threats. We also note that the end-to-end SPC generally occupies much less time  than one timeslot. Since the time taken for one channel use is inversely proportional to the available bandwidth, we have $\delta_i[n] = \frac{l_i[n]}{W}~\forall n$, where $i\in\{u,~d\}$, and $\delta_{i}[n]$  indicates the time needed for a finite blocklength SPC in either the uplink or downlink transmission.

The three-dimensional Cartesian coordinates of Alice, Bob, and Eve are denoted by $\q_a=[x_a,y_a,0]^T$, $\q_b=[x_b,y_b,0]^T$, and $\q_e=[x_e,y_e,0]^T$, respectively, where $[\cdot]^T$ represents the transpose operator. Note that all the communication nodes are assumed to be a part of the same network; thus, the location of the communication nodes is perfectly known. Nonetheless, the information exchanged between Alice and Bob should be kept secret from Eve for confidentiality purposes. We assume that the UR's flight time horizon is set to $T$, and is split into $N$ timeslots such that  $T= N \delta_t$. Since the SPC duration is small,  we assume that the UR's location over the transmission phase in each timeslot remains approximately unchanged, but varies from one timeslot to another.  Furthermore, the  altitude of the UR is fixed at $H$, which typically corresponds to the minimum flying altitude for being avoided by obstacles or mountainous areas, establishing LoS-dominant channel components \cite{TatarMamaghani2021}. Therefore, the UR's location in timeslot $n$  can be denoted by $\q_r[n]=[x[n], y[n], H]^T$. As such, the UR's continuous trajectory can be approximated by $(N+1)$ waypoints, i.e., $\{\q_r[n]\}^{N}_{n=1}$. Assuming that the UR's initial and final locations are denoted by $\q_i=[x_i, y_i, H]^T$ and $\q_f=[x_f, y_f, H]^T$, respectively, the following mobility constraints are  imposed on the UR trajectory:
\begin{subequations}
\begin{align}\label{const:mob}
&\hspace{-2mm}\cst{1}:~\q_r[1] = \q_i,~ \q_r[N] = \q_f,\\
&\hspace{-2mm}\cst{2}:~\|\q_r[n+1]-\q_r[n] \|\leq v^{max} \delta_t,~n=1,\cdots, N\hspace{-1mm}-\hspace{-1mm}1
\end{align}
\end{subequations}
where the constraint \cst{2} limits the displacement of the UR for consecutive timeslots, and $v^{max}$ indicates the maximum flight velocity of the UR. 

The AG channels are assumed to be dominated by path-loss with negligible fading \cite{TatarMamaghani2020}. Thus, for the Alice-UR link, the UR-Eve link, and the UR-Bob link, denoted as $h_{ra}[n]$, $h_{re}[n]$, and $h_{rb}[n]~\forall n$, respectively,  we express their LoS-dominant channel power gains as
\begin{align}
    h_{rj}[n]=\frac{\beta_0}{\|\q_r[n] - \q_j\|^2},~ \forall n, ~ j \in\{a, e, b\}
\end{align}
where $\beta_0$ denotes the path-loss at a reference distance under omnidirectional propagation. Furthermore, since both Alice and Eve are terrestrial nodes, the channel model for the Alice-Eve link constitutes both distance-dependent attenuation and small-scale Rayleigh fading \cite{TatarMamaghani2018}, the power gain of which can be represented as
\begin{align}
    h_{ae}[n]= \frac{\beta_0}{\|\q_a - \q_e\|^\alpha}\zeta[n],~\forall n
\end{align}
where $\zeta[n]$ is a unit-mean exponential random variable, and $\alpha$ is the corresponding environmental path-loss exponent, with a typical range between $2 < \alpha \leq 4$.

Delay tolerance in SPC is crucial as the communication system must deliver sensitive information quickly to be effective. Thus, for the considered short-packet delay-sensitive system, the requirement on delay tolerance can be imposed by constraining the number of total blocklengths per transmission to be less than the maximum allowable end-to-end delay, expressed as
\begin{align}
    &\cst{3}:~\sum_{i} l_i[n] \leq L^{max},~l_i[n] \in \mathbb{Z}^{0+},~ i\in\{u, d\},~ \forall n
\end{align}
where $L^{max}$ denotes the maximum latency tolerance and $ \mathbb{Z}^{0+}$ represents the set of nonnegative integers.

\vspace*{5mm}
\section{Problem Formulation}

Let $\varepsilon$ and $\eta$ denote the network node's decoding error probability and  information leakage, respectively. For simplicity, we also assume that the additive white Gaussian noise (AWGN) power at any node is equal for all timeslots and denoted by $\sigma^2$. Consequently, we express the achievable average SR in bits per channel use  for the short-packet uplink transmission in timeslot $n$, according to \cite{Yang2019}, as
\begin{align}\label{secrate_1}
    \tilde{R}^{sec}_u[n] &= \E_{\zeta[n]}\Big\{\log_2\left(\frac{1+\gamma_r[n]}{1+\gamma_{ae}[n]}\right)- \sqrt{\frac{V(\gamma_r[n])}{l_u[n]}}\Qi\left( \varepsilon\right) \nonumber \\
    &- \sqrt{\frac{V(\gamma_{ae}[n])}{l_u[n]}}\Qi\left( \eta\right) \Big\},~\forall n\\
    &\stackrel{(a)}{\approx}  \log_2\left(\frac{1+\gamma_r[n]}{1+\bar{\gamma}_{ae}[n]}\right)-
\sqrt{\frac{V(\gamma_r[n])}{l_u[n]}}\Qi\left( \varepsilon\right) 
\nonumber\\
&- \sqrt{\frac{V(\bar{\gamma}_{ae}[n])}{l_u[n]}}\Qi\left( \eta\right)\treq {R}^{sec}_u[n],~ \forall n\label{approx_sruplink}
\end{align}
where $\E_x\{\cdot\}$ indicates expectation over the random variable $x$, and $\Qi(x)$ is the inverse of the complementary Gaussian cumulative distribution function $Q(x)$, defined as $Q(x)=\int^{\infty}_{x}\frac{1}{\sqrt{2\pi}}\e^{-\frac{r^2}{2}}dr$.  Moreover, $\gamma_r[n]$ and $\gamma_{ae}[n]$, denoting the received signal-to-noise ratios (SNRs) at the UR and Eve in timeslot $n$, are given respectively by
\begin{align}
\gamma_r[n] &=  \frac{\rho_{a}}{\|\q_r[n] - \q_a\|^2},~\forall n\\
\gamma_{ae}[n]&= \frac{\rho_{a}}{\|\q_a - \q_e\|^\alpha}{\zeta}[n],~\forall n
\end{align}
Note that the approximation $(a)$ in \eqref{approx_sruplink} follows from Jensen's inequality, and  $\bar{\gamma}_{ae}[n] =  \frac{\rho_{a}}{\|\q_a - \q_e\|^\alpha}$, where  $\rho_{a} = \frac{p_a\beta_0}{\sigma^2}$. Furthermore, the function $V(\cdot)$ indicates channel dispersion, which can be mathematically expressed, according to  \cite{Yang2019},  as
\begin{align}
   V(\gamma) = \log^2_2(\e)\left[1-\left(1+\gamma\right)^{-2}\right],~ \forall n
\end{align}
Note that $V(\gamma)$ is a monotonically increasing  function of the SNR $\gamma$. 

Likewise, the achievable  SR for short-packet downlink transmission in timeslot $n$ is given by
\begin{align}\label{secrate_2}
     \begin{split} 
    R^{sec}_d[n] =   \log_2&\left(\frac{1+\gamma_b[n]}{1+\gamma_{re}[n]}\right)- \sqrt{\frac{ V(\gamma_b[n])}{l_d[n]}}\Qi\left( \varepsilon\right) \\
     &- \sqrt{\frac{ V(\gamma_{re}[n])}{l_d[n]}}\Qi\left( \eta\right),~ \forall n
       \end{split} 
\end{align}
where $\gamma_b[n]$ and $\gamma_{re}[n]$ denote  the received SNRs at Bob and Eve in timeslot $n$, given respectively by
\begin{align}
\gamma_b[n] &=  \frac{\rho_r}{\|\q_r[n] - \q_b\|^2},~\forall n\\
\gamma_{re}[n]&= \frac{\rho_{r}}{\|\q_r[n]- \q_e\|^2},~\forall n
\end{align}
with $\rho_r = \frac{p_r\beta_0}{\sigma^2}$. Now, considering that Alice and the UR securely encode the transmit short-packet data in timeslot $n$ to maintain the desired reliability and security requirements of the considered system $(\varepsilon,  \eta)$, we define the secrecy throughput metric as the rate of the effective number of securely transmitted information bits in bits per second (bps) as
\begin{align}
\hspace{-3mm} \bar{B}_s[n] =\frac{1- \varepsilon}{\delta_t}~\Big[\min(R^{sec}_u[n]l_u[n] , R^{sec}_d[n]l_d[n])\Big]_+,~\forall n
\end{align}
 where $[x]_+=\max(x, 0)$.

Our objective is to optimize the secrecy performance of the proposed UAV-SPC relaying system by designing the transmission blocklengths $\L=\{l_u[n],~l_d[n],~\forall n\}$, and the UR trajectory $\Q=\{\q[n], ~\forall n\}$. The resulting optimization maximizes the Effective Average Secrecy Throughput (\east) over the mission duration, which is formulated as
\begin{align}\label{opt_prob}
\Prob{}:& \stackrel{}{\underset{\{\L, \Q\}}{\mathrm{max}}}~~ \east = \frac{1}{N}\sum_{n=1}^{N} \bar{B}_s[n]\nonumber\\
&\quad\text{s.t.}\qquad \cst{1}-\cst{3}.
\end{align}
Note that \Prob{} is a nonconvex optimization problem due to the nonconvex objective function with nonsmooth operator $[\cdot]_+$, and highly coupled optimization variables. Thus, it is too challenging to be solved optimally. First, we note that the nonsmoothness of the objective function in \Prob{} can be handled since at the optimal point, $\bar{B}_s[n]$ should hold a nonnegative value. Otherwise, by setting $l_u[n]=0$ and/or $l_d[n]=0$ (i.e., no transmission) in the given timeslot, one obtains $\bar{B}_s[n]=0$, which violates the optimality. In light of this, we  remove the nonsmoothness operator from the objective function without impacting the optimal solution. Here, introducing a slack variable vector $\pmb{\tau}=\{\tau[n], \forall n\}$, we convert \Prob{} to a more tractable version, whose objective function is differentiable and serves as a lower bound on that of the original problem:
\begin{subequations}
\begin{align}\label{opt_pro_alt}
\Prob{1}:& \stackrel{}{\underset{\{\L,~\Q,~\pmb{\tau}\}}{\mathrm{max}}}~~~\frac{1-\varepsilon}{T}\sum_{n=1}^{N} \tau[n] \nonumber\\
&\quad\text{s.t.}\qquad\cst{1}-\cst{3},\\
&\qquad\qquad  {R}^{sec}_u[n] l_u[n] \geq \tau[n],~\forall n\label{15b}\\
&\qquad\qquad  {R}^{sec}_d[n] l_d[[n] \geq \tau[n],~\forall n\label{15c}
\end{align}
\end{subequations}
In the following, we propose a low-complexity iterative solution to solve \Prob{1}  based on the BSCA algorithm, wherein we optimize each block of variables while keeping others unchanged in an alternating manner. Such an algorithm  generally approaches a locally optimal solution but with guaranteed convergence.

\vspace*{5mm}
\section {Proposed Solution}\label{sec:solution}
In this section,  we divide \Prob{1} into two subproblems: i) short-packet blocklength optimization, and ii) UR trajectory optimization. Thereafter, we propose an efficient overall iterative algorithm. Note that in the sequel, we omit the fixed multiplicative term $\frac{1-\varepsilon}{T}$ from the objective function, which has no impact on the proposed solution.

\vspace*{5mm}
\subsection{Short-packet Blocklength Optimization}
In this subsection, we optimize the blocklength vectors of both the uplink and downlink SPC, i.e., $\L=\{l_u[n], l_d[n],~\forall n\}$, while keeping other variables fixed. As such,  the corresponding subproblem can be written as
\begin{align}
\Prob{2}:& \stackrel{}{\underset{\L,~\pmb{\tau} }{\mathrm{max}}}~~~\sum_{n=1}^{N} \tau[n]\nonumber\\
&\quad\text{s.t.}\quad\cst{3},~\eqref{15b},~\eqref{15c}.
\end{align}
Note that \Prob{2} is a nonlinear integer programming, which is in general a challenging NP-hard  problem. Nevertheless, analyzing the objective function of \Prob{2}, we see that it is non-decreasing with respect to $\L$, implying that the constraint \cst{3} should be satisfied with equality at the optimal point, i.e., $l_u[n] + l_d[n]= L^{max}~\forall n$; otherwise, increasing the blocklength leads to an increase in the objective function, which violates the optimality. Hence, we can solve \Prob{2} via a simple 1D search over the discrete set $\mathcal{L}=\{0, 1, 2 \cdots, L^{max}\}$ to obtain the optimal blocklengths for the uplink transmissions:
\begin{align}\label{blocklengthOptim}
l^{opt}_u[n]=\argmax_{x\in\mathcal{L}} \bar{B}_s[n]\Big|_{l_u[n]=x, l_d[n]=L^{max}-x},~\forall n
\end{align}
Then, the optimal blocklengths for the downlink transmissions can be determined by 
\[l^{opt}_d[n] = L^{max} - l^{opt}_u[n]~\forall n.\]

\vspace*{5mm}
\subsection{UR Trajectory Optimization}
This subsection explores the joint optimization of the UR's motion and altitude. In light of this, we recast \Prob{1} to optimize $\Q$, while keeping the other variables fixed, which gives
\begin{subequations}
\begin{align}\label{trj_subprob}
\Prob{3}:& \stackrel{}{\underset{\{\q_r,~\pmb{\tau}\}}{\mathrm{max}}}~~~\sum_{n=1}^{N} \tau[n] \nonumber\\
&\quad\text{s.t.}~~~~~ \cst{1}-\cst{2},\\
\begin{split}
    &\log_2(1+\gamma_r[n]) - b_0\sqrt{1-(1+\gamma_r[n])^{-2}} \\
    &\quad\geq b_2 \tau[n] + b_1,~ \forall n \label{p4.cst2}
\end{split}\\
\begin{split}
    &\log_2\left(\frac{1+\gamma_b[n]}{1+{\gamma}_{re}[n]}\right) - c_0\sqrt{1-(1+\gamma_b[n])^{-2}} \\
    &\quad- c_1\sqrt{1-(1+{\gamma}_{re}[n])^{-2}} \geq c_2 \tau[n],~ \forall n\label{p4.cst3}
\end{split}
\end{align}
\end{subequations}
where \[b_0 = \frac{\Qi(\varepsilon)\log_2 \e}{\sqrt{l_u[n]}},~b_2 = \frac{1}{l_u[n](1-\varepsilon)},\] 
\[b_1 = \log_2({1+{\gamma}_{ae}[n]}) - \sqrt{\frac{V({\gamma}_{ae}[n])}{l_u[n]}}\Qi( \eta),\] 
\[c_0 = \frac{\Qi(\varepsilon)\log_2 \e}{\sqrt{l_d[n]}},~c_1 = \frac{\Qi(\eta)\log_2 \e}{\sqrt{l_d[n]}},~c_2 = \frac{1}{l_d[n](1-\varepsilon)}.\]
We stress that \Prob{3} is still a nonconvex optimization problem due to nonconvex constraints \eqref{p4.cst2} and \eqref{p4.cst3}. In the following, we focus on transforming these constraints into convex approximates to make the problem tractable.

\subsubsection{Convex reformulation of \eqref{p4.cst2}} We equivalently write \eqref{p4.cst2} in a more tractable way by introducing nonnegative slack variables $\pmb{\lambda}=\{\lambda_1[n], \lambda_2[n],~\forall n\}$ and $\pmb{\beta}=\{\beta_1[n],~\forall n\}$, as
\begin{subequations}
\begin{align}
    &\log_2(1+\lambda_1[n]) - b_0\beta_1[n]  \geq b_2 \tau[n] + b_1,~\forall n\label{ineq1}\\
    & {\rho_a}\lambda_2[n] \geq {\|\q_r[n] - \q_a \|^2},~\forall n\label{ineq3}\\
    &\lambda_1[n] \lambda_2[n] \leq 1,~\forall n\label{ineq2}\\  
    &\beta^2_1[n] \geq 1-(1+\lambda_1[n])^{-2},~\forall n \label{ineq4}
\end{align}
\end{subequations}
We note that the constraints \eqref{ineq1} and \eqref{ineq3} are convex, while the additional constraints \eqref{ineq2} and \eqref{ineq4}, introduced to ensure the smoothness of \Prob{3}, are nonconvex. We stress that \eqref{ineq3}-\eqref{ineq4} should hold with equality at the optimal point. Before proceeding further, we present a lemma below. 
\begin{lemma}\label{lemma_3}
    Let $f(x,y)=\frac{1}{xy}$ with $x, y > 0$. At any given point $(x_0, y_0)$ in the domain of $f$, the following function serves as a global lower bound on $f(x,y)$ \cite{Boyd2006}, i.e., 
    \begin{align}\label{upperbound}
        f_{lb}(x,y;x_0,y_0) =  -\frac{x\,y_{0}+x_{0}\,y-3\,x_{0}\,y_{0}}{{x_{0}}^2\,{y_{0}}^2} \leq f(x,y).
    \end{align}
\end{lemma}
Thus, at a given point $(\q^{lo}_r,\pmb{\lambda}^{lo},\pmb{\beta}^{lo})$, the convex approximations of the constraints \eqref{ineq2} and \eqref{ineq4} are
\begin{subequations}\label{cst_cvx}
\begin{align} 
 & 1\leq  
  f_{lb}(\lambda_1[n],\lambda_2[n];\lambda^{lo}_1[n],\lambda^{lo}_2[n]),~\forall n\label{ineq2_cvx}\\
     &\ln(\beta_1[n]) + \ln(1+\lambda_1[n]) \geq  g(\lambda_1[n]; \lambda^{lo}_1[n]),~\forall n \label{ineq5_cvx}
\end{align}
\end{subequations}
where the convex function $g(x; x_0)$ is defined as
\[
    g(x; x_0)=A_0(x_0)+ A_1(x_0)\left(x-x_0\right),
\]
with $A_0(x)$ and $A_1(x)$ being defined for $x>0$ as
\[A_0(x)=\frac{1}{2} \ln \left(x\left[2+x\right]\right)\]
and
\[A_1(x)=\frac{x+1}{x \left(x+ 2\right)},\]
respectively. Here, \eqref{ineq2_cvx} follows from Lemma \ref{lemma_3} and \eqref{ineq5_cvx} follows from the concavity of the logarithm function.

\subsubsection{Convex reformulation of \eqref{p4.cst3}}
Introducing the nonnegative  slack variables $\pmb{\omega}=\{\omega_1[n], \omega_2[n],~\forall n\}$, $\pmb{\psi}=\{\psi_1[n],~\forall n\}$, $\mathbf{u}=\{u_1[n],~\forall n\}$, and $\mathbf{v}=\{v_1[n], v_2[n],~\forall n\}$, we reformulate \eqref{p4.cst3} into approximate convex constraints, using Lemma \ref{lemma_3} and \cite[Lemma 3]{TatarMamaghani2021}, as
 \begin{subequations} \label{p4.cst3_cvx}
 \begin{align}
\begin{split}
        &\log_2\left({1+\omega_1[n]}\right) -  \log_2\left(1+{u^{-1}_1[n]}\right)  \\
        &\qquad\geq  c_0\psi_1[n] + c_1v_1[n] +c_2 \tau[n],~\forall n
\end{split}\\
& 1\leq   f_{lb}(\omega_1[n],\omega_2[n];\omega^{lo}_1[n],\omega^{lo}_2[n]),~\forall n\\
    & \rho_r\omega_2[n] \geq \|\q_r[n] - \q_b \|^2,~\forall n\\
     &\ln(\psi_1[n]) + \ln(1+\omega_1[n]) \geq   g(\omega_1[n]; \omega^{lo}_1[n]),~\forall n   \\
  & \rho_r u_1[n] \leq  2\left({\q^{lo}_r[n]} - {\q}_e\right)^T\q_r[n] + d^{lo}_0[n],~\forall n\\
     &  \ln(v_1[n]) + \ln(1+v_2[n]) \geq   g(v_2[n]; v^{lo}_2[n]),~\forall n \\
& 
  u_1[n] \geq v^{-1}_2[n],~\forall n  
\end{align}
\end{subequations}
where $d^{lo}_0[n] = \|{\q}_e\|^2-\|{\q^{lo}_r[n]}\|^2$ and $(\q^{lo}_r,\pmb{\omega}^{lo},\pmb{\psi}^{lo}, \mathbf{v}^{lo}, \mathbf{u}^{lo})$ is the given local point.  We now express the convex reformulation of subproblem \Prob{3} as
\begin{align}\label{trj_subprob_cvx}
\Prob{3.1}:& \stackrel{}{\underset{\{\q_r, \pmb{\tau}, \pmb{\lambda}, \pmb{\beta}, \pmb{\omega}, \pmb{\psi}, \mathbf{u},\mathbf{v}\}}{\mathrm{max}}}~~~\sum_{n=1}^{N} \tau[n] \nonumber\\
&\quad\text{s.t.}~\cst{1}-\cst{2}, \eqref{ineq1}, \eqref{ineq3}, \eqref{cst_cvx}, \eqref{p4.cst3_cvx}.
\end{align}
Since \Prob{3.1} is convex, it can be efficiently solved by standard convex optimization tools such as CVX \cite{CVXResearch2012}.

 \vspace*{5mm} 
\subsection{Overall Iterative Algorithm}

In this subsection, we propose an overall iterative algorithm based on the sequential block optimization summarized in Algorithm \ref{myalgo1}. \newline

\removelatexerror
  \begin{algorithm}[H]{
  \caption{Overall EAST optimization algorithm.}\label{myalgo1}
  1:~\textbf{Initialize:}~
  Iteration index $i=0$, choose a feasible local point (${\L}^{(i)},  \Q^{(i)}$);\\
  2:~\textbf{Repeat:} \\
  2.1:~Calculate $\east^{(i)}$, update $i \gets i+1$;\\
  2.2:~Solve \Prob{2} via \eqref{blocklengthOptim}, then update ${\L}^{(i)}=\{{l}^{(i)}_u[n], {l}^{(i)}_d[n],~\forall n\}$;\\
  2.3:~Given ${\L}^{(i)}$, initialize slack variables $\left(\pmb{\lambda}^{lo}, \pmb{\beta}^{lo}, \pmb{\omega}^{lo}, \pmb{\psi}^{lo}, \mathbf{u}^{lo}, \mathbf{v}^{lo}\right)$, and solve \Prob{3.1}, then update $\Q^{(i)}=\{\q^{(i)}_r[n]= [x^{(i)}_r[n], y^{(i)}_r[n], H]^T,~\forall n\}$;\\
  2.4:~ Calculate $\east^{(i)}$ at the new point $\left({\L}^{(i)},  \Q^{(i)}\right)$;\\
  3:~\textbf{Until:} $\|\east^{(i)} - \east^{(i-1)}\| \leq \epsilon$;}\\
  \end{algorithm}
\vspace*{5mm}

It can be proved that the proposed algorithm is guaranteed to converge to a local optimum commencing from a feasible point. Moreover, the time complexity of Algorithm \ref{myalgo1}, based on the complexity of each convex subproblem and the convergence accuracy parameter $\epsilon$, can be approximately obtained as $\O\left(N^{3.5}\log_2(\frac{1}{\epsilon})\right)$, where $\O(\cdot)$ represents big-O notation.  The overall complexity order of Algorithm \ref{myalgo1} is polynomial, and thus our proposed approach can be reasonably implemented for energy-limited UAV-IoT scenarios with SPC.

\vspace*{5mm}
\section{Numerical Results and Discussion}\label{sec:numerical}

In this section, we demonstrate  the EAST enhancement achieved by our proposed optimization algorithm  for the considered UAV-aided SPC-IoT scenario. To
exhibit  the effectiveness of our joint trajectory and blocklength design in Algorithm \ref{myalgo1}, labeled~\prop, we compare it with the following benchmark schemes:
\begin{itemize} 
\item Trajectory Design with Fixed Blocklength (\td): With fixed coding blocklengths, only the UR trajectory is optimized using \Prob{3.1} in a sequential manner.  
\item Blocklength Design with Fixed Trajectory (\rd): Keeping the UR trajectory fixed, only the optimization of blocklengths  is taken via solving  \Prob{2} using \eqref{blocklengthOptim}.
\end{itemize}
Table \ref{table:sysParams} lists the system parameters used in this section, unless otherwise stated. These parameters are mainly adopted in the literature such as in \cite{Feng2021} and \cite{TatarMamaghani2021}. In addition, the initial feasible trajectory of the UR, $\Q^{(0)}$, is taken to be on a direct line with fixed speed from the initial location to the final location. The uplink and downlink transmission blocklengths per timeslot are initialized as $l^{(0)}_u[n]=l^{(0)}_d[n]=\frac{L^{max}}{2}~\forall n$. 


\begin{table}[t]
\caption{System parameters}
\centering
\resizebox{\columnwidth}{!}{%
\begin{tabular}{l c} 
\midrule \midrule 
\textbf{Simulation parameter (notation)} & \textbf{Value}\\ [0.5ex] 
\midrule 
Transmit power ($p_a$)& $20$ dBm \\
Relaying power ($p_r$)& $20$ dBm \\
Reference channel power gain ($\beta_0$) & $-70$ dB\\
Terrestrial path-loss exponent ($\alpha$) & $3$\\
Channel noise power ($\sigma^2$) & $-140$ dBm\\
Transmission period ($\delta_t$) & $1$ s\\
Maximum latency tolerance ($L^{max}$) & 400\\
Mission time ($T$) & $100$ s\\
UAV's altitude ($H$) & $60$ m\\
Maximum flying speed ($v^{max}$) & $30$ m/s\\
UAV's initial location ($\q_i$) & $[-500, -1000, 60]^T$ m\\
UAV's final location ($\q_f$) & $[1000, 500, 60]^T$ m\\
Alice's location ($\q_a$) & $[-700,0, 0]^T$ m\\
Bob's location ($\q_b$) & $[700,0, 0]^T$ m\\
Eve's location (${\q}_e$) & $[-500,900, 0]^T$ m\\
Decoding error probability ($\varepsilon$)& $10^{-3}$\\
Information leakage parameter ($\eta$) & $10^{-2}$\\
Convergence threshold  ($\epsilon$) & $10^{-3}$\\
\midrule 
\end{tabular}}
\label{table:sysParams}
\end{table}

\begin{figure}[t]
        \centering
        \includegraphics[width= \columnwidth]{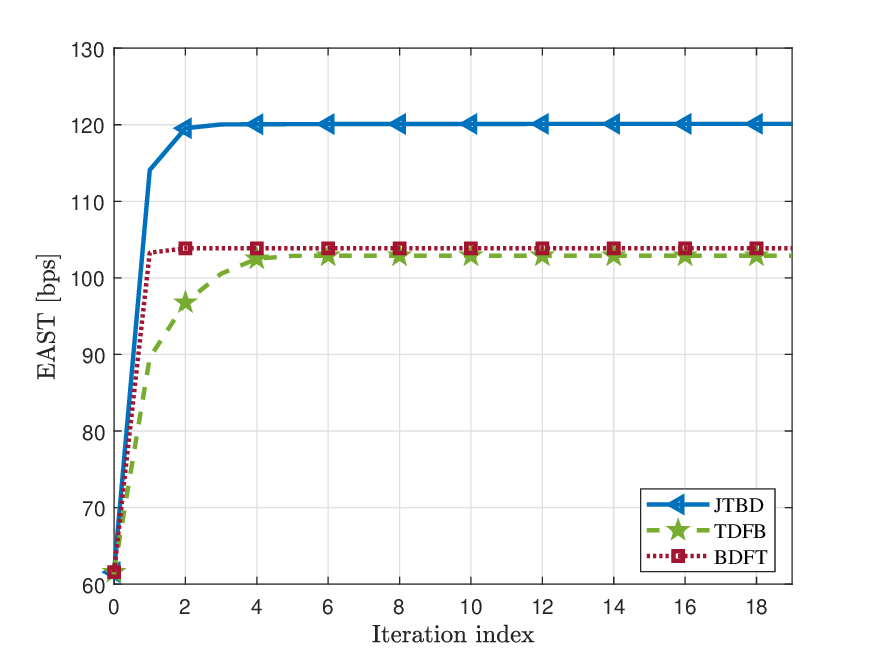}
        \caption{EAST vs. iteration index.}
        \label{sim:fig2east}
\end{figure}

Fig. \ref{sim:fig2east} depicts the EAST performance against the iteration index for all schemes to verify the quick convergence of Algorithm \ref{myalgo1} and  the validity of our analysis, as well as to demonstrate the performance advantage of our joint design. We observe from the figure that the EAST is non-decreasing over the iteration index for  all algorithms, and that convergence occurs quickly in just a few iterations. Our proposed \prop~approach achieves the best EAST performance amongst all. For example, \prop~can reach up to $120$ bps, approximately $14\%$ more than  both the \rd~and~\td~designs, and nearly twice the EAST of the initial feasible setting. Also, Fig. \ref{sim:fig2east} shows that using the baseline trajectory and optimizing transmission blocklengths is more important for SPC than optimizing the trajectory with fixed blocklengths, while the joint design of both is clearly preferable concerning the EAST performance metric. 

\begin{figure}[!t]
  \centering
    \begin{subfigure}[t]{\columnwidth}
        \centering
        \includegraphics[width= \linewidth]{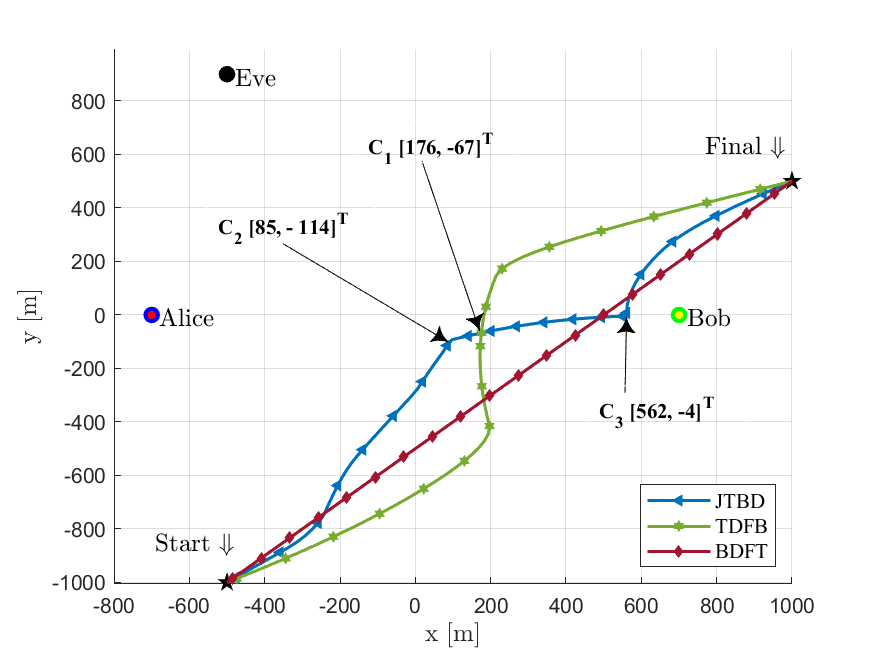}
        \caption{UR's trajectory profile.}
        \label{sim:fig2traj}
    \end{subfigure}
~
 \begin{subfigure}[t]{\columnwidth}
        \centering
        \includegraphics[width= \linewidth]{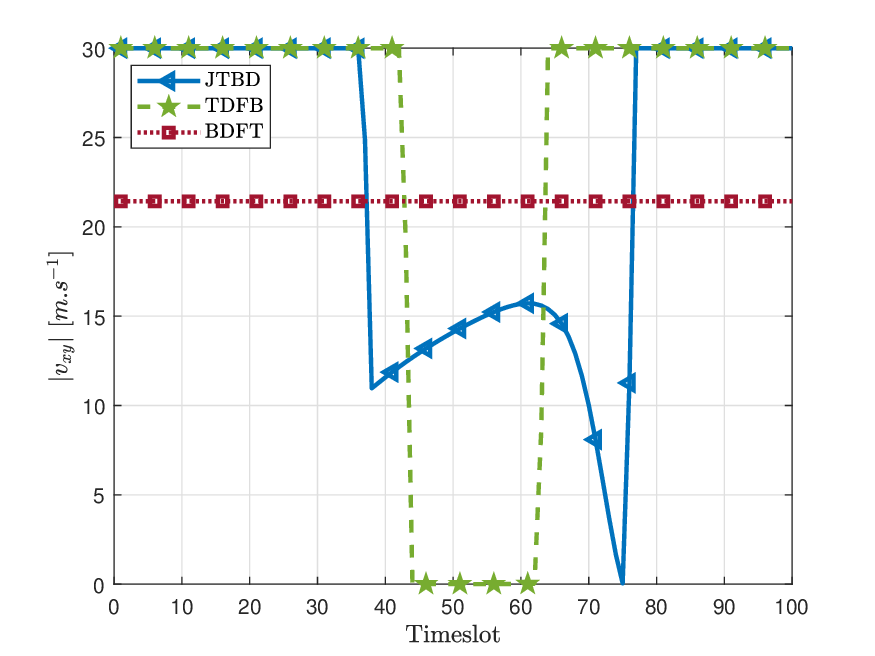}
        \caption{UR's velocity vs. timeslot.}
        \label{sim:fig2vel}
\end{subfigure}
\\
\begin{subfigure}[t]{\columnwidth}
    \centering
    \includegraphics[width=\linewidth]{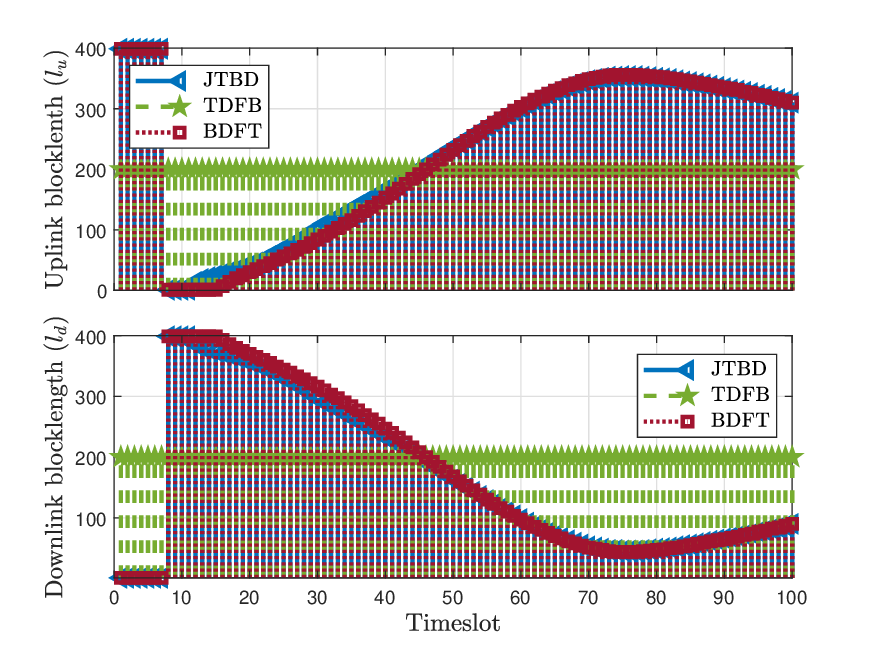}
    \caption{Blocklengths vs. timeslot.}
    \label{sim:fig2blocklength}
\end{subfigure}

     \caption{Designed UR's trajectory and velocity profiles, and coding blocklengths according to different schemes.} 
      \label{sim:fig2}
\end{figure}

Fig. \ref{sim:fig2} illustrates the trajectory and velocity profiles of the UR, as well as both the uplink and downlink short-packet blocklengths for the specific realization of the user locations based on different design scenarios.  We observe from Figs. \ref{sim:fig2}(\subref{sim:fig2traj}) and \ref{sim:fig2}(\subref{sim:fig2vel}) that in contrast to the other schemes, for \td~the UR attempts to fly with the maximum velocity while heading towards a position between Alice and Bob located at the coordinate $\mathbf{C}_1 = [176, -67, 60]^T$ m, and hovering at that point as long as possible, i.e., between $T=44$~s and $T=62$~s. This solution greatly improves the EAST compared with the initial direct-path trajectory with fixed velocity, but not as much as \rd~and our proposed~\prop. We also observe from Fig. \ref{sim:fig2}(\subref{sim:fig2blocklength}) that when the UR is farther from Bob,  larger downlink coding blocklengths are adopted, and they reduce in length as the UR approaches Bob. As the UR flies away from Alice, the proposed algorithm efficiently increases the uplink blocklength, which ultimately enhances the EAST. Interestingly, the blocklength designs of both \prop~and \rd~follow a very similar trend, even though their corresponding trajectories are different. Nevertheless, when both the trajectory design and blocklength optimization are taken into account as in the proposed \prop~design, the UR demonstrates effective navigation and ability to considerably enhance the EAST compared to other benchmarks. Specifically, the effective trajectory for the \prop~scheme requires the UR to fly with full speed from the initial location $\q_i$ to the position marked as $\mathbf{C}_2=[85, -114,60]^T$ m, according to the path illustrated in Fig. \ref{sim:fig2}(\subref{sim:fig2traj}). Then, a sharp velocity drop occurs at $T=36$~s, enabling the UR to move with relatively low speed from the position $\mathbf{C}_2$ to a location with the coordinate $\mathbf{C}_3=[562, -4, 60]^T$ m, for improving the overall EAST performance. Finally, the UR travels through an arc path tilted away from Bob while maximally increasing its velocity at timeslot $T=75$~s and maintaining this velocity so that the last part of the mission from  $\mathbf{C}_3$ to the final location $\q_f$ could be accomplished by the end of the specified duration. Overall, this observation reinforces the significance of communication blocklength and trajectory co-design for the secrecy performance improvement of aerial relaying with SPC.

\vspace*{5mm}
\section{Conclusion}\label{sec:conclusion}
This work presented the design of a secure and reliable  UAV-IoT relaying system with SPC. To optimize the EAST performance of the system, an effective joint design approach that incorporates the UAV trajectory and both uplink and downlink blocklengths was proposed, and was shown to achieve quick convergence with low complexity. The efficacy of the proposed approach was evaluated through numerical simulations, and the results demonstrated the superiority of our \prop~scheme in terms of EAST compared to benchmarks that only consider either trajectory design or blocklength optimization. Our results further indicated that  both the uplink and downlink blocklengths should be adaptively adjusted 
according to the UAV location along the trajectory.

\newpage
\bibliographystyle{IEEEtran}
\bibliography{myReferences}

\begin{thebibliography}{10}
\providecommand{\url}[1]{#1}
\csname url@samestyle\endcsname
\providecommand{\newblock}{\relax}
\providecommand{\bibinfo}[2]{#2}
\providecommand{\BIBentrySTDinterwordspacing}{\spaceskip=0pt\relax}
\providecommand{\BIBentryALTinterwordstretchfactor}{4}
\providecommand{\BIBentryALTinterwordspacing}{\spaceskip=\fontdimen2\font plus
\BIBentryALTinterwordstretchfactor\fontdimen3\font minus \fontdimen4\font\relax}
\providecommand{\BIBforeignlanguage}[2]{{%
\expandafter\ifx\csname l@#1\endcsname\relax
\typeout{** WARNING: IEEEtran.bst: No hyphenation pattern has been}%
\typeout{** loaded for the language `#1'. Using the pattern for}%
\typeout{** the default language instead.}%
\else
\language=\csname l@#1\endcsname
\fi
#2}}
\providecommand{\BIBdecl}{\relax}
\BIBdecl

\bibitem{Durisi2016}
G.~Durisi, T.~Koch, and P.~Popovski, ``{Toward massive, ultrareliable, and low-latency wireless communication with short packets},'' \emph{Proc. {IEEE}}, vol. 104, no.~9, pp. 1711--1726, Sept. 2016.

\bibitem{Feng2021}
C.~Feng and H.~M. Wang, ``{Secure short-packet communications at the physical layer for 5G and beyond},'' \emph{IEEE Commun. Stand. Mag.}, vol.~5, no.~3, pp. 96--102, Sept. 2021.

\bibitem{Shirvanimoghaddam2019}
{M. Shirvanimoghaddam \textit{et al.}}, ``Short block-length codes for ultra-reliable low latency communications,'' \emph{IEEE Commun. Mag.}, vol.~57, pp. 130--137, Feb. 2019.

\bibitem{Yang2015}
N.~Yang, L.~Wang, G.~Geraci, M.~Elkashlan, J.~Yuan, and M.~Di~Renzo, ``{Safeguarding 5G wireless communication networks using physical layer security},'' \emph{IEEE Commun. Mag.}, vol.~53, pp. 20--27, Apr. 2015.

\bibitem{TatarMamaghani2018}
M.~Tatar~Mamaghani, A.~Kuhestani, and K.-K. Wong, ``{Secure two-way transmission via wireless-powered untrusted relay and external jammer},'' \emph{IEEE Trans. Veh. Technol.}, vol.~67, pp. 8451--8465, Sept. 2018.

\bibitem{TatarMamaghani2020}
M.~Tatar~Mamaghani and Y.~Hong, ``{Improving PHY-security of UAV-enabled transmission with wireless energy harvesting: Robust trajectory design and communications resource allocation},'' \emph{IEEE Trans. Veh. Technol.}, vol.~69, pp. 8586--8600, Aug. 2020.

\bibitem{Poor2017}
H.~V. Poor and R.~F. Schaefer, ``{Wireless physical layer security},'' \emph{Proc. Natl. Acad. Sci. U.S.A.}, vol. 114, no.~1, pp. 19--26, Jan. 2017.

\bibitem{Yang2019}
W.~Yang, R.~F. Schaefer, and H.~V. Poor, ``{Wiretap channels: Nonasymptotic fundamental limits},'' \emph{IEEE Trans. Inf. Theory}, vol.~65, pp. 4069--4093, July 2019.

\bibitem{Zheng2020}
T.-X. Zheng, H.-M. Wang, D.~W.~K. Ng, and J.~Yuan, ``{Physical-layer security in the finite blocklength regime over fading channels},'' \emph{IEEE Trans. Wirel. Commun.}, vol.~19, no.~5, pp. 3405--3420, May 2020.

\bibitem{Feng2022}
C.~Feng, H.~M. Wang, and H.~V. Poor, ``{Reliable and secure short-packet communications},'' \emph{IEEE Trans. Wirel. Commun.}, vol.~21, pp. 1913--1926, Mar. 2022.

\bibitem{Wang2019e}
H.-M. Wang, Q.~Yang, Z.~Ding, and H.~V. Poor, ``{Secure short-packet communications for mission-critical IoT applications},'' \emph{IEEE Trans. Wirel. Commun.}, vol.~18, pp. 2565--2578, May 2019.

\bibitem{Wu2021}
{Q. Wu \textit{et al.}}, ``{A comprehensive overview on 5G-and-beyond networks with UAVs: From communications to sensing and intelligence},'' \emph{IEEE J. Sel. Areas Commun.}, vol.~39, pp. 2912--2945, Oct. 2021.

\bibitem{Wang2019}
H.~M. Wang, X.~Zhang, and J.~C. Jiang, ``{UAV-involved wireless physical-layer secure communications: Overview and research directions},'' \emph{IEEE Wirel. Commun.}, vol.~26, no.~5, pp. 32--39, Oct. 2019.

\bibitem{TatarMamaghani2021}
M.~Tatar~Mamaghani and Y.~Hong, ``{Joint trajectory and power allocation design for secure artificial noise aided UAV communications},'' \emph{IEEE Trans. Veh. Technol.}, vol.~70, pp. 2850--2855, Mar. 2021.

\bibitem{Boyd2006}
S.~Boyd and L.~Vandenberghe, \emph{{Convex Optimization}}.\hskip 1em plus 0.5em minus 0.4em\relax Cambridge, U.K.: Cambridge Univ. Press, 2004.

\bibitem{CVXResearch2012}
\BIBentryALTinterwordspacing
M.~Grant and S.~Boyd, ``{CVX: Matlab Software for Disciplined Convex Programming, Version 2.2},'' Jan. 2020. [Online]. Available: \url{http://cvxr.com/cvx}
\BIBentrySTDinterwordspacing

\end{thebibliography}

\end{document}